\documentclass[12pt]{article}
\usepackage{a4wide,amsmath,amssymb,alltt,graphicx}
\newcommand\accel{_{\mathrm{accel}}}
\newcommand\cores{_{\mathrm{cores}}}
\newcommand{\eg}{e.g.\ }
\newcommand{\ie}{i.e.\ }
\newcommand\Code[1]{\ensuremath{\texttt{#1}}}
\newcommand\Var[1]{\ensuremath{\mathit{#1}}}
\newcommand\ul[1]{\underline{\vphantom{g}#1}}
\newcommand\Vol{\mathop{\mathrm{Vol}}}

\makeatletter
\def\reportno#1{\gdef\@reportno{#1}}
\def\@maketitle{%
  \hfill{\small\begin{tabular}[t]{r}%
    \@reportno
  \end{tabular}\par}%
  \vskip 2em%
  \begin{center}%
    \let \footnote \thanks
    {\large \@title \par}%
    \vskip 1.5em%
    {
      \lineskip .5em%
      \begin{tabular}[t]{c}%
        \@author  
      \end{tabular}\par}%
    \vskip 1em%
    {
     \@date}%
  \end{center}%
  \par
  \vskip 1.5em}
\makeatother

\begin{document}

\reportno{MPP--2014--327 \\
PACS: 02.60.Jh}

\author{T. Hahn \\
Max-Planck-Institut f\"ur Physik \\
F\"ohringer Ring 6, D--80805 Munich, Germany}

\title{Concurrent Cuba}

\maketitle

\begin{abstract}
The parallel version of the multidimensional numerical integration 
package Cuba is presented and achievable speed-ups discussed.
\end{abstract}


\section{Introduction}

Cuba is a library for multidimensional numerical integration written in 
C99 with interfaces for Fortran, C/C++, and Mathematica.  Cuba offers a 
choice of four independent routines for multidimensional numerical 
integration, Vegas, Suave, Divonne, and Cuhre, with very different 
characteristics \cite{cuba,cuba-parallel,cuba-checkpointing}.

Numerical integration is perfectly suited for parallel execution, which 
can significantly speed up the computation as it generally incurs only a 
very small overhead.  Several features for concurrent sampling were 
added in Cuba versions 3 and 4, for both parallelization and 
vectorization.  The objective was to make parallel use of Cuba 
very easy, ideally automatic, and led to the following design decisions:

1. No kind of Message Passing Interface is used, as that requires extra 
software to be installed.  That is, the parallelization is restricted to 
one computer, using operating-system functions only.  A standard setup 
these days is a single CPU with a number of cores, say 4 or 8.  
Utilizing many more compute nodes, as one could potentially do with MPI, 
is more of a theoretical option anyway since the speed-ups cannot be 
expected to grow linearly, see Sect.\ \ref{sect:perf} on Performance.

2. Cuba uses \Code{fork}/\Code{wait} rather than the \Code{pthread*} 
functions.  The latter are slightly more efficient because parent and 
child share their memory space, but for the same reason they also 
require a reentrant integrand function, and the programmer may not have 
control over reentrancy in all languages (\eg Fortran's I/O is typically 
non-reentrant).  \Code{fork} on the other hand creates a completely 
independent copy of the running process and thus works for any integrand 
function.

3. Changing the number of cores to use should not require a re-compile, 
in particular as the program image should be able to run on several 
computers (with possibly different numbers of cores) simultaneously. 
This is solved through an environment variable \Code{CUBACORES}, which
defaults to the number of idle cores on the present system.

The present note describes the concurrent features of Cuba 4 and 
discusses their performance.  For installation and usage details not 
related to parallelization, the user is referred to the Cuba manual.


\section{Parallelization model}

Cuba uses a master--worker model.  The master process orchestrates the 
parallelization but does not count towards the number of cores, \eg 
\Code{CUBACORES = 4} means four workers and one master.  Very 
importantly, the samples are generated by the master process only and 
distributed to the workers, such that random numbers are never used more 
than once.


\subsection{`Simple' parallelization}

The parallelization of Cuba naturally focusses on the main sampling 
routine \Code{DoSample}, called by all Cuba integrators, which has been 
abstracted since the very first Cuba version simply because it is 
implemented very differently in Fortran/C/C++ and in Mathematica. 
\Code{DoSample} in principle parallelizes straightforwardly on $N$ 
cores:
\begin{quote}
\begin{tabbing}
Serial version:\qquad
	\= sample $n$ points. \\[1ex]
Parallel version:
	\> sample $n/N$ points on core 1, \\[-.5ex]
	\> \quad $\vdots$ \\[-.3ex]
	\> sample $n/N$ points on core $N$.
\end{tabbing}
\end{quote}
The actual distribution strategy is somewhat more involved and is
described in Sect.~\ref{sect:cores}.

One appealing aspect of the parallelization through \Code{DoSample} is 
that changes are made at a conceptually `low' level of the integration, 
\ie parallelization can be dealt with without modifying the integration 
algorithm itself.


\subsection{Parallelization in Divonne}
\label{sect:divonne}

The speed-ups achieved with Divonne by parallelizing \Code{DoSample} 
alone were generally unsatisfactory and significantly below those of the 
other integrators, \eg $\lesssim$ 1.5 on four cores.

The Divonne algorithm works in three phases:
\begin{itemize}
\item Partitioning Phase: Split the integration region into subregions 
with approximately equal spread, defined as
$$
\mathop{\mathrm{Spread}}(r) = \frac 12 \Vol(r)
  \Bigl(\sup_{\vec x\in r} f(\vec x\,) -
        \inf_{\vec x\in r} f(\vec x\,)\Bigr),
$$
where the minimum and maximum of the integrand $f$ are sought using 
methods from numerical optimization.

\item Sampling Phase: Sample the subregions independently with the same 
number of points each.  That number is extrapolated from the results of 
the Partitioning Phase.

\item Refinement Phase: Further subdivide or sample again if results 
from the Partitioning and Sampling Phase do not agree within their 
estimated errors.
\end{itemize}
It turned out that the Partitioning Phase was crucial for attaining 
reasonable speed-ups and needed special treatment: Firstly, the original 
partitioning algorithm divided the regions recursively (and with a 
minimum recursion depth, too) and had to be `un-recursed', mainly by 
better bookkeeping of the subregions.

Secondly, the Partitioning Phase was modified such that each core 
receives an entire region to subdivide, not just a list of points (as 
\Code{DoSample} does).  In particular the minimum/maximum search, during 
which only one point at a time is sampled, is distributed much more 
efficiently this way.  The other two phases were not so critical 
precisely because they sample more points per region.

By moving the parallelization one level `up' as it were, \ie no longer 
at the `lowest' (sampling) level of the integration, the genuine Divonne 
algorithm becomes more entwined with the parallelization, of course, and 
also the master--worker communication becomes more complex.


\section{User Guide}
\label{sect:user}

The parallelization procedure is rather different in Fortran/C/C++ and 
in Mathematica.  We shall deal with the latter first because it needs 
only a short explanation.  The remainder of this chapter is then devoted 
to the Fortran/C/C++ case.


\subsection{Parallelization in Mathematica}

The Mathematica version of Cuba performs its sampling through a function 
\Code{MapSample}.  By default this is identical to \Code{Map}, \ie the 
serial version, so to parallelize one merely needs to redefine 
\Code{MapSample = ParallelMap} (after loading Cuba).

If the integrand depends on user-defined symbols or functions, their 
definitions must be distributed to the workers beforehand using 
\Code{DistributeDefinitions} and likewise required packages must be 
loaded with \Code{ParallelNeeds} instead of \Code{Needs}; this is 
explained in detail in the Mathematica manual.


\subsection{Parallelization in Fortran and C/C++}

In Fortran and C/C++ the Cuba library can (and usually does) 
automatically parallelize the sampling of the integrand.  It 
parallelizes through \Code{fork} and \Code{wait} which, though slightly 
less performant than pthreads, do not require reentrant code.  
(Reentrancy may not even be under full control of the programmer, for 
example Fortran's I/O is usually non-reentrant.)  Worker processes are 
started and shut down only as few times as possible, however, so the 
performance penalty is really quite minor even for non-native fork 
implementations such as Cygwin's.  Parallelization is not available on 
native Windows for lack of the \Code{fork} function.

The communication of samples to and from the workers happens through IPC 
shared memory (\Code{shmget} and colleagues), or if that is not 
available, through a \Code{socketpair} (two-way pipe).  Remarkably, the 
former's anticipated performance advantage turned out to be hardly 
perceptible.  Possibly there are cache-coherence issues introduced by 
several workers writing simultaneously to the same shared-memory area.


\subsubsection{Invocation in Fortran}
\label{sect:commonargs}

For reference, the prototypes of the integrators in Cuba 4 are repeated 
here; their detailed description is left to the Cuba manual.  Only the 
underlined arguments are relevant for the following discussion.
\begin{alltt}
  subroutine vegas(ndim, ncomp, integrand, userdata, \ul{nvec},
    epsrel, epsabs, flags, seed, mineval, maxeval,
    nstart, nincrease, nbatch, gridno, statefile, \ul{spin},
    neval, fail, integral, error, prob)
\end{alltt}
\begin{alltt}
  subroutine suave(ndim, ncomp, integrand, userdata, \ul{nvec},
    epsrel, epsabs, flags, seed, mineval, maxeval,
    nnew, flatness, statefile, \ul{spin},
    nregions, neval, fail, integral, error, prob)
\end{alltt}
\begin{alltt}
  subroutine divonne(ndim, ncomp, integrand, userdata, \ul{nvec},
    epsrel, epsabs, flags, seed, mineval, maxeval,
    key1, key2, key3, maxpass,
    border, maxchisq, mindeviation,
    ngiven, ldxgiven, xgiven, nextra, peakfinder,
    statefile, \ul{spin},
    nregions, neval, fail, integral, error, prob)
\end{alltt}
\begin{alltt}
  subroutine cuhre(ndim, ncomp, integrand, userdata, \ul{nvec},
    epsrel, epsabs, flags, mineval, maxeval,
    key, statefile, \ul{spin},
    nregions, neval, fail, integral, error, prob)
\end{alltt}
The external function which computes the integrand is expected to be
declared as
\begin{alltt}
  integer function integrand(ndim, x, ncomp, f, userdata, \ul{nvec}, \ul{core}, ...)
  integer ndim, ncomp, nvec, core
  double precision x(ndim,nvec), f(ncomp,nvec)
\end{alltt}
The integrand receives \Code{nvec} \Code{ndim}-dimensional samples in 
\Code{x} and is supposed to fill the array \Code{f} with the 
corresponding \Code{ncomp}-component integrand values.  The return value 
is irrelevant unless it is $-999$, in the case of which the integration 
will be aborted immediately.  Note that \Code{nvec} indicates the actual 
number of points passed to the integrand here and may be smaller than 
the \Code{nvec} given to the integrator.

The dots represent optional arguments provided by Vegas, Suave, and 
Divonne (see manual).  Also \Code{userdata}, \Code{nvec}, and 
\Code{core} are optional and may be omitted if unused, \ie as in former 
Cuba versions the integrand may minimally be declared (for $\Code{nvec} 
= 1$) as
\begin{verbatim}
  integer function integrand(ndim, x, ncomp, f)
  integer ndim, ncomp
  double precision x(ndim), f(ncomp)
\end{verbatim}


\subsubsection{Invocation in C/C++}

The C/C++ prototypes are contained in \Code{cuba.h}.  They are 
reproduced here for reference.  Again, only the underlined arguments are 
relevant for the present discussion.
\begin{verbatim}
typedef int (*integrand_t)(const int *ndim, const double x[],
  const int *ncomp, double f[], void *userdata);

typedef void (*peakfinder_t)(const int *ndim, const double b[],
  int *n, double x[]);
\end{verbatim}
\begin{alltt}
void Vegas(const int ndim, const int ncomp,
  integrand_t integrand, void *userdata, \ul{const int nvec},
  const double epsrel, const double epsabs,
  const int flags, const int seed,
  const int mineval, const int maxeval,
  const int nstart, const int nincrease, const int nbatch,
  const int gridno, const char *statefile, \ul{void *spin},
  int *neval, int *fail,
  double integral[], double error[], double prob[])
\end{alltt}
\begin{alltt}
void Suave(const int ndim, const int ncomp,
  integrand_t integrand, void *userdata, \ul{const int nvec},
  const double epsrel, const double epsabs,
  const int flags, const int seed,
  const int mineval, const int maxeval,
  const int nnew, const double flatness,
  const char *statefile, \ul{void *spin},
  int *nregions, int *neval, int *fail,
  double integral[], double error[], double prob[])
\end{alltt}
\begin{alltt}
void Divonne(const int ndim, const int ncomp,
  integrand_t integrand, void *userdata, \ul{const int nvec},
  const double epsrel, const double epsabs,
  const int flags, const int seed,
  const int mineval, const int maxeval,
  const int key1, const int key2, const int key3,
  const int maxpass, const double border,
  const double maxchisq, const double mindeviation,
  const int ngiven, const int ldxgiven, double xgiven[],
  const int nextra, peakfinder_t peakfinder,
  const char *statefile, \ul{void *spin},
  int *nregions, int *neval, int *fail,
  double integral[], double error[], double prob[])
\end{alltt}
\begin{alltt}
void Cuhre(const int ndim, const int ncomp,
  integrand_t integrand, void *userdata, \ul{const int nvec},
  const double epsrel, const double epsabs,
  const int flags,
  const int mineval, const int maxeval,
  const int key, const char *statefile, \ul{void *spin},
  int *nregions, int *neval, int *fail,
  double integral[], double error[], double prob[])
\end{alltt}
The \verb=integrand_t= type intentionally declares only a minimalistic 
integrand type (and even the \Code{userdata} argument could be omitted 
further).  A more complete declaration is
\begin{alltt}
typedef int (*integrand_t)(const int *ndim, const double x[],
  const int *ncomp, double f[], void *userdata,
  \ul{const int *nvec}, \ul{const int *core}, ...);
\end{alltt}
where the dots stand for extra arguments passed by Vegas, Suave, and 
Divonne (see manual) not needed in the following.  In the presence of an 
\Code{nvec} argument, \Code{x} and \Code{f} are actually two-dimensional 
arrays, \Code{x[*nvec][*ndim]} and \Code{f[*nvec][*ncomp]}.

The integrand receives \Code{*nvec} \Code{*ndim}-dimensional samples in 
\Code{x} and is supposed to fill the array \Code{f} with the 
corresponding \Code{*ncomp}-component integrand values.  The return 
value is irrelevant unless it is $-999$, which signals immediate 
abortion of the integration.  Note that \Code{*nvec} indicates the 
actual number of points passed to the integrand here and may be smaller 
than the \Code{*nvec} given to the integrator.


\subsubsection{Starting and stopping the workers}
\label{sect:spinning}

The workers are usually started and stopped automatically by Cuba's 
integration routines, but the user may choose to start them manually or 
keep them running after one integration and shut them down later, \eg at 
the end of the program, which can be slightly more efficient.  The 
latter mode is referred to as `Spinning Cores' and must be employed with 
certain care, for running workers will not `see' subsequent changes in 
the main program's data (\eg global variables, common blocks) or code 
(\eg via \Code{dlsym}) unless special arrangements are made (\eg shared 
memory).

The spinning cores are controlled through the `\Code{spin}' argument of 
the Cuba integration routines (Sect.~\ref{sect:commonargs}):
\begin{itemize}
\item A value of \Code{-1} or \Code{\%VAL(0)} (in Fortran) or 
\Code{NULL} (in C/C++) tells the integrator to start and shut down the 
workers autonomously.  This is the usual case.  No workers will still be 
running after the integrator returns.  No special precautions need to be 
taken to communicate \eg global data to the workers.  Note that it is 
expressly allowed to pass a `naive' \Code{-1} (which is an 
\Code{integer}, not an \Code{integer*8}) in Fortran.

\item Passing a zero-initialized variable for \Code{spin} instructs the 
integrator to start the workers but keep them running on return and 
store the `spinning cores' pointer in \Code{spin} for future use.  The 
spinning cores must later be terminated explicitly by \Code{cubawait}, 
thus invocation would schematically look like this:

\hfill\begin{minipage}{.4\hsize}
\begin{verbatim}
integer*8 spin
spin = 0
call vegas(..., spin, ...)
...
call cubawait(spin)
\end{verbatim}
\end{minipage}\hfill\begin{minipage}{.4\hsize}
\begin{verbatim}
void *spin = NULL;

Vegas(..., &spin, ...);
...
cubawait(&spin);
\end{verbatim}
\end{minipage}\hfill

\item A non-zero \Code{spin} variable is assumed to contain a valid 
`spinning cores' pointer either from a former integration or an explicit 
invocation of \Code{cubafork}, as in:

\hfill\begin{minipage}{.4\hsize}
\begin{verbatim}
integer*8 spin
call cubafork(spin)
call vegas(..., spin, ...)
...
call cubawait(spin)
\end{verbatim}
\end{minipage}\hfill\begin{minipage}{.4\hsize}
\begin{verbatim}
void *spin;
cubafork(&spin);
Vegas(..., &spin, ...);
...
cubawait(&spin);
\end{verbatim}
\end{minipage}\hfill
\end{itemize}


\subsubsection{Accelerators and Cores}
\label{sect:cores}

Based on the strategy used to distribute samples, Cuba distinguishes two 
kinds of workers.

Workers of the first kind are referred to as `Accelerators' even though 
Cuba does not actually send anything to a GPU or Accelerator in the 
system by itself -- this can only be done by the integrand routine.  The 
assumption behind this strategy is that the integrand evaluation is 
running on a device so highly parallel that the sampling time is more or 
less independent of the number of points, up to the number of threads 
$p\accel$ available in hardware.  Cuba tries to send exactly $p\accel$ 
points to each core -- never more, less only for the last batch.  To 
sample \eg 2400 points on three accelerators with $p\accel = 1000$, Cuba 
sends batches of 1000/1000/400 and not, for example, 800/800/800 or 
1200/1200.  The number of accelerators $n\accel$ and their value of 
$p\accel$ can be set through the environment variables
\begin{tabbing}
\verb|   CUBAACCEL=|$n\accel$ \hspace{10em}\= (default: 0) \\
\verb|   CUBAACCELMAX=|$p\accel$    \> (default: 1000)
\end{tabbing}
or, superseding the environment, an explicit
\begin{alltt}
   call cubaaccel(\(n\accel\), \(p\accel\))
\end{alltt}

CPU-bound workers are just called `Cores'.  Their distribution strategy 
is different in that all available cores are used and points are 
distributed evenly.  In the example above, the batches would be 
800/800/800 thus.  Each core receives at least 10 points, or else fewer 
cores are used.  If no more than 10 points are requested in total, Cuba 
uses no workers at all but lets the master sample those few points.  
This happens during the partitioning phase of Divonne, for instance, 
where only single points are evaluated in the minimum/maximum search. 
Conversely, if the division of points by cores does not come out even, 
the remaining few points ($< n\cores$) are simply added to the existing 
batches, to avoid an extra batch because of rounding.  Sampling 2001 
points on two cores with $p\cores = 1000$ will hence give two 
batches 1001/1000 and not three batches 1000/1000/1.

Although there is typically no hardware limit, a maximum number of 
points per core, $p\cores$, can be prescribed for Cores, too.  Unless 
the integrand is known to evaluate equally fast at all points, a 
moderate number for $p\cores$ (10000, say) may actually increase 
performance because it effectively load-levels the sampling.  For, a 
batch always goes to the next free core so it doesn't matter much 
if one core is tied up with a batch that takes longer.

The number of cores $n\cores$ and the value of $p\cores$ can be set 
analogously through the environment variables
\begin{tabbing}
\verb|   CUBACORES=|$n\cores$ \hspace{10em}\= (default: no.\ of idle cores) \\
\verb|   CUBACORESMAX=|$p\cores$ \> (default: 10000)
\end{tabbing}
If \Code{CUBACORES} is unset, the idle cores on the present system are 
taken (total cores minus load average), which means that a program 
calling a Cuba routine will by default automatically parallelize on the 
available cores.  Again, the environment can be overruled with an 
explicit
\begin{alltt}
   call cubacores(\(n\cores\), \(p\cores\))   
\end{alltt}
Using the environment has the advantage, though, that changing the 
number of cores to use does not require a re-compile, which is 
particularly useful if one wants to run the program on several computers 
(with potentially different numbers of cores) simultaneously, say in a 
batch queue.

The integrand function may use the `\Code{core}' argument 
(Sect.~\ref{sect:commonargs}) to distinguish Accelerators ($\Code{core} 
< 0$) and Cores ($\Code{core}\geqslant 0$).  The special value 
$\Code{core} = 32768$ ($2^{15}$) indicates that the master itself is 
doing the sampling.


\subsubsection{Worker initialization}

User subroutines for (de)initialization may be registered with
\begin{alltt}
   call cubainit(initfun, initarg)           \textrm{Fortran}
   call cubaexit(exitfun, exitarg)
\end{alltt}
\begin{alltt}
   cubainit(initfun, initarg);               \textrm{C/C++}
   cubaexit(exitfun, exitarg);
\end{alltt}
and will be executed in every process before and after sampling.
Passing a null pointer (\Code{\%VAL(0)} in Fortran, \Code{NULL} in
C/C++) as the first argument unregisters either subroutine.

The init/exit functions are actually called as
\begin{alltt}
   call initfun(initarg, core)               \textrm{Fortran}
   call exitfun(exitarg, core)
\end{alltt}
\begin{alltt}
   initfun(initarg, &core);                  \textrm{C/C++}
   exitfun(exitarg, &core);
\end{alltt}
where \Code{initarg} and \Code{exitarg} are the user arguments given 
with the registration (arbitrary in Fortran, \Code{void *} in C/C++) 
and \Code{core} indicates the core the function is being executed on, 
with (as before) $\Code{core} < 0$ for Accelerators, 
$\Code{core}\geqslant 0$ for Cores, and $\Code{core} = 32768$ for the 
master.

On worker processes, the functions are respectively executed after 
\Code{fork} and before \Code{wait}, independently of whether the worker 
actually receives any samples.  The master executes them only when 
actual sampling is done.
For Accelerators, the init and exit functions are typically used to set 
up the device for the integrand evaluations, which for many devices must 
be done per process, \ie after the \Code{fork}.


\subsubsection{Concurrency issues}

By creating a new process image, \Code{fork} circumvents all memory 
concurrency, to wit: each worker modifies only its own copy of the 
parent's memory and never overwrites any other's data.  The programmer 
should be aware of a few potential problems nevertheless:
\begin{itemize}
\item Communicating back results other than the intended output from the 
integrand to the main program is not straightforward because, by the 
same token, a worker cannot overwrite any common data of the master, it 
will only modify its own copy.

Data exchange between workers is likewise not directly possible.  For 
example, if one worker stores an intermediate result in a common block, 
this will not be seen by the other workers.

Possible solutions include using shared memory (\Code{shmget} etc., see 
App.~\ref{app:shm}) and writing the output to file (but see next item 
below).

\item \Code{fork} does not guard against competing use of other common 
resources.  For example, if the integrand function writes to a file 
(debug output, say), there is no telling in which order the lines will 
end up in the file, or even if they will end up as complete lines at 
all.  Buffered output should be avoided at the very least; better still, 
every worker should write the output to its own file, \eg with a 
filename that includes the process id, as in:
\begin{verbatim}
   character*32 filename
   integer pid
   data pid /0/
   if( pid .eq. 0 ) then  
     pid = getpid()
     write(filename,'("output.",I5.5)') pid
     open(unit=4711, file=filename)
   endif
\end{verbatim}

\item Fortran users are advised to flush (or close) any open files 
before calling Cuba, \ie \Code{call flush(\Var{unit})}.  The reason is 
that the child processes inherit all file buffers, and \emph{each} of 
them will write out the buffer content at exit.  Cuba preemptively 
flushes the system buffers already (\Code{fflush(NULL)}) but has no 
control over Fortran's buffers.
\end{itemize}
For debugging, or if a malfunction due to concurrency issues is 
suspected, a program should be tested in serial mode first, \eg by 
setting $\Code{CUBACORES} = 0$ (Sect.~\ref{sect:cores}).


\subsubsection{Vectorization}

Vectorization means evaluating the integrand function for several points 
at once.  This is also known as Single Instruction Multiple Data (SIMD) 
paradigm and is different from ordinary parallelization where 
independent threads are executed concurrently.  It is usually possible 
to employ vectorization on top of parallelization.

Vector instructions are commonly available in hardware, \eg on x86 
platforms under acronyms such as SSE or AVX.  Language support varies: 
Fortran 90's syntax naturally embeds vector operations.  Many C/C++ 
compilers offer auto-vectorization options, some have extensions for 
vector data types (usually for a limited set of mathematical functions), 
and even hardware-specific access to the CPU's vector instructions.  And 
then there are vectorized libraries of numerical functions available.

Cuba cannot automatically vectorize the integrand function, of course, 
but it does pass (up to) \Code{nvec} points per integrand call 
(Sect.~\ref{sect:commonargs}).  This value need not correspond to the 
hardware vector length -- computing several points in one call can also 
make sense \eg if the computations have significant intermediate results 
in common.  The actual number of points passed is indicated through the 
corresponding \Code{nvec} argument of the integrand.

\medskip

A note for disambiguation: The \Code{nbatch} argument of Vegas is 
related in purpose but not identical to \Code{nvec}.  It internally 
partitions the sampling done by Vegas but has no bearing on the number 
of points given to the integrand.  On the other hand, it it pointless to 
choose $\Code{nvec} > \Code{nbatch}$ for Vegas.


\section{Performance}

\subsection{Test setup}

Parallelization entails a certain overhead as usual, so the efficiency
will depend on the `cost' of an integrand evaluation, \ie the more
`expensive' (time-consuming) it is to sample the integrand, the better
the speed-up will be.  

The timing measurements in the following figures were made with the 11 
integrands of the demo code included in the Cuba distribution, which 
were originally chosen to highlight different aspects of the 
integrators.  These are simple one-liners and for timing purposes 
`infinitely' fast.  To tune the cost of the integrands, a calibrated 
delay loop was inserted into the integrand functions.

The calibration was necessary because the system time resolution was 
visibly imprinted on the first versions of the plots.  In a separate 
program, the delay loop was executed with no upper bound but with the 
timer set to interrupt at 10 seconds.  The number of cycles performed 
per second was recorded in a file and in subsequent timing measurements 
was used to compute the upper bound of the delay loop.  This guaranteed 
that, for a prescribed delay, each integrand was slowed down by the same 
number of delay-loop turns, with a $\mu$sec precision of the delay.

The number of repetitions of each integration was moreover adjusted for 
each integrator to make the serial version run for 240 sec.  That leaves 
at least 30 sec run-time per invocation even for the ideal speed-up of 8 
(on an i7), long enough to make time-measurement errors negligible.

To exclude systematic effects, all measurements were done on the same 
i7-2600K Linux machine, idle except for the timing program.  The i7 
processor has four real and eight virtual (hyperthreaded) cores, \ie 
eight register sets but only four arithmetic units.  Beyond four cores, 
indicated by a line in the plots, CPU effects are thus expected on top 
of Cuba's `pure' scaling behavior.


\subsection{Timing measurements}
\label{sect:perf}

\begin{figure}[p]
\centerline{\includegraphics[width=.9\hsize]{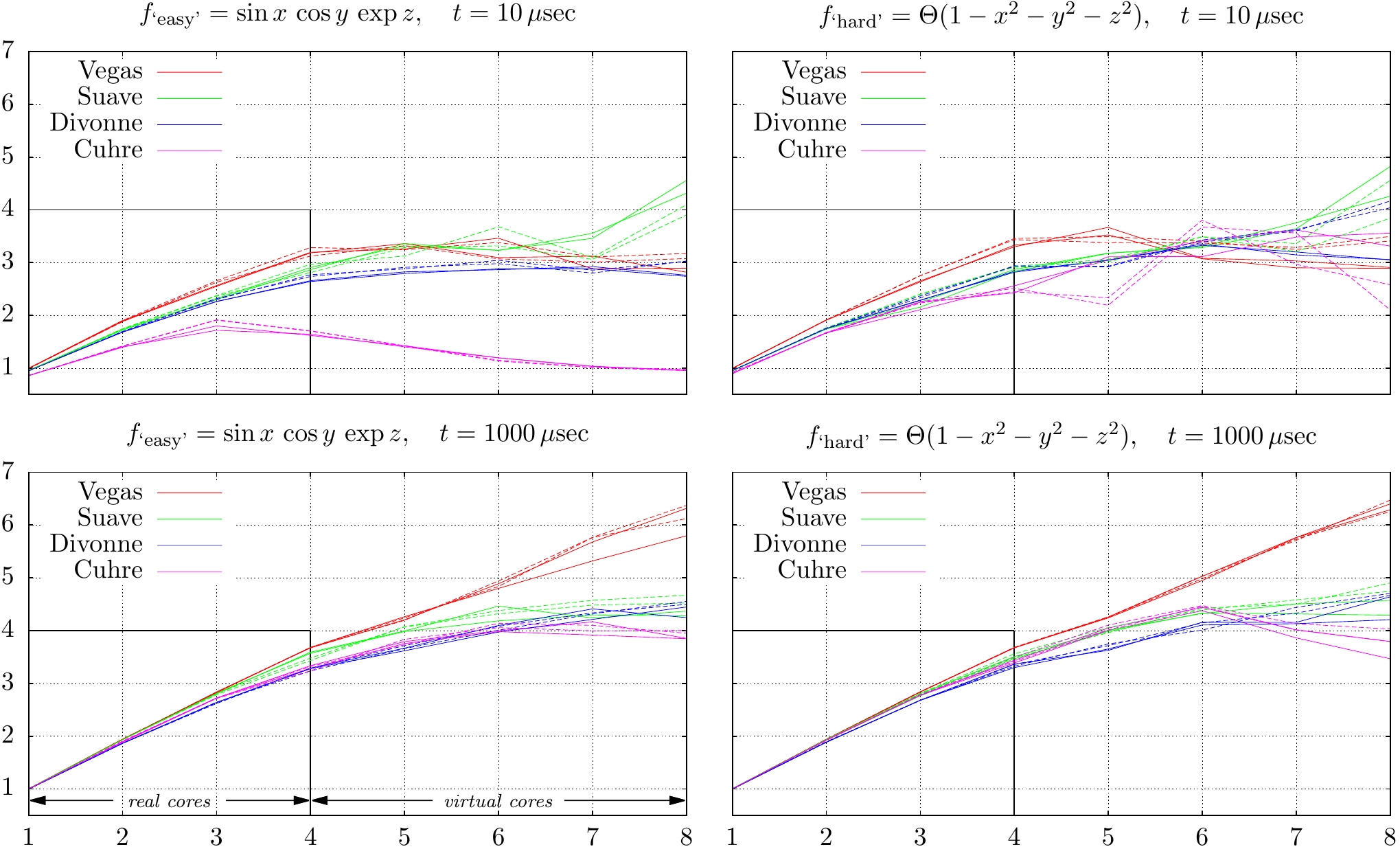}}
\caption{\label{fig:easyhard}Cuba speed-ups for three-dimensional 
integrals with $10^{-4}$ requested accuracy.
\textit{Left:} `easy' integrand,
\textit{right:} `hard' integrand.
\textit{Top:} `fast' integrand ($10\,\mu$sec),
\textit{bottom:} `slow' integrand ($1000\,\mu$sec per evaluation).
\textit{Solid:} shared memory,
\textit{dashed:} socketpair communication
(two curves each to show fluctuations in timing).}
\end{figure}

Fig.\ \ref{fig:easyhard} shows the speed-ups for an `easy' and a `hard' 
one of the 11 integrands of the demo program included in the Cuba 
package for two different integrand delays.  Also in the one-core case 
the parallel version was deployed (one master, one worker), which 
explains why the timings normalized to the serial version are below 1, 
in the top row visibly so.

The first, expected, observation is that parallelization is worthwhile 
only for not-too-fast integrands.  This is not a major showstopper, 
however, as many interesting integrands fall into this category anyway.  
What appears to be a drastic underperformance of Cuhre in the `easy' 
case can in fact be attributed to Cuhre's outstanding efficiency: it 
delivers a result correct to almost all digits with around 300 samples.  
In such a case, Cuba may for efficiency choose not to fill all available 
cores and relative to the full number of cores this shows up as a 
degradation.

\begin{figure}[p]
\centerline{\includegraphics[width=.9\hsize]{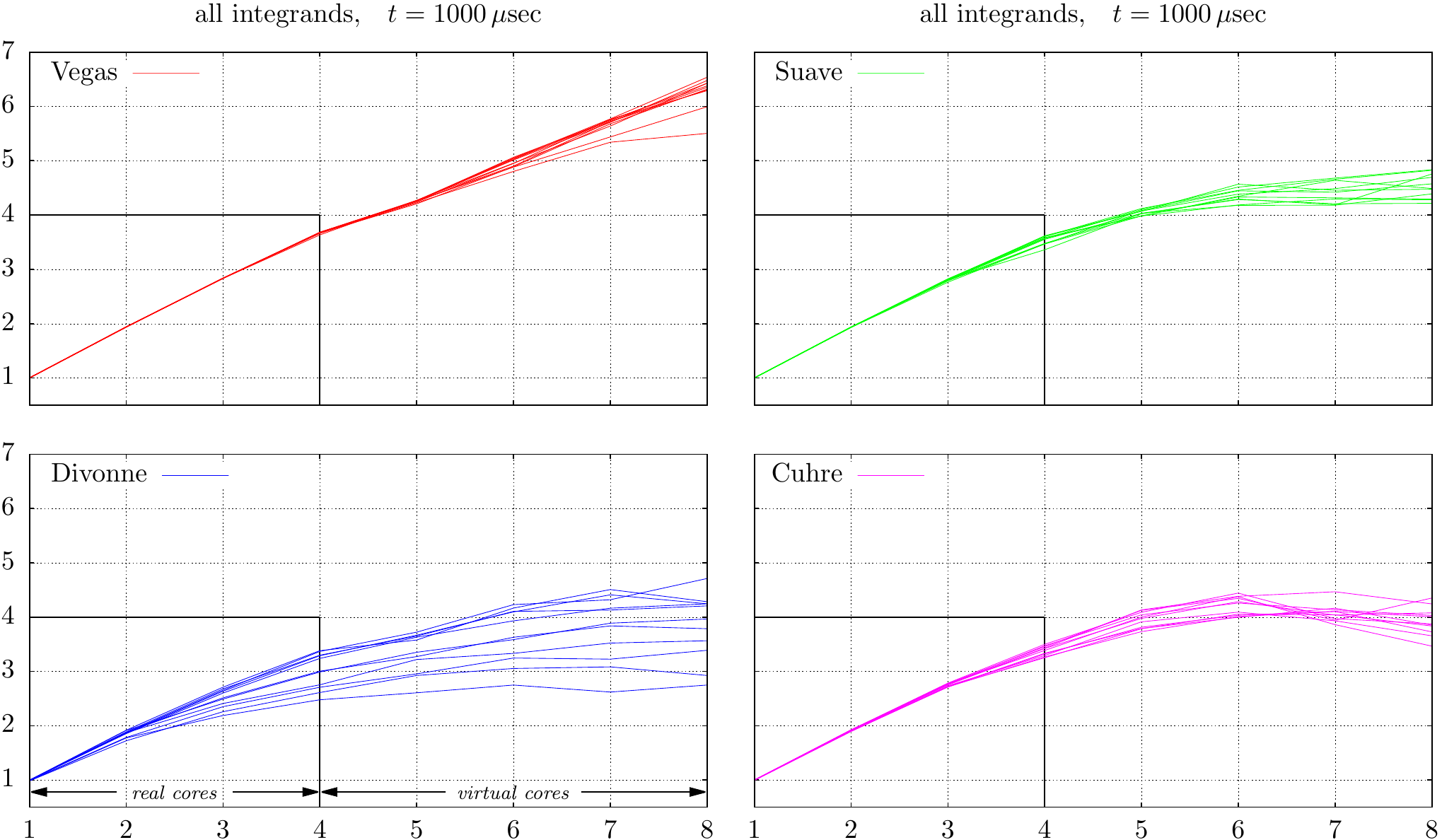}}
\caption{\label{fig:allintegrands}Comparison of the parallelization 
efficiency for all 11 integrands.}
\end{figure}

The second observation is that parallelization works best for 
`simple-minded' integrators, \eg Vegas.  This is showcased even better 
in Fig.~\ref{fig:allintegrands}.  The `intelligent' algorithms are 
generally much harder to parallelize because they don't just do 
mechanical sampling but take into account intermediate results, make 
extra checks on the integrand (\eg try to find extrema), etc.  This is 
particularly true for Divonne, see Sect.\ \ref{sect:divonne}.  Then 
again, the `intelligent' algorithms are usually faster to start with,
\ie converge with fewer points sampled, which compensates for the lack 
of parallelizability.

\section{Summary}

The Cuba library for multidimensional numerical integration now features 
concurrent sampling and thereby achieves significant speed-ups.  No 
extra software needs to be installed since only operating-system 
functions are used.  No reentrancy is required for the integrand 
function since \Code{fork}/\Code{wait} is applied.  Parallelization is 
usually switched on automatically but can be controlled through API 
calls or the environment.

Cuba is available from \Code{http://feynarts.de/cuba} and licensed under 
the LGPL.  The download contains a manual which gives full details on 
installation and usage.


\section*{Acknowledgments}

The author thanks Alexander Smirnov for valuable suggestions and 
comments and the MIAPP Workshop `Challenges, Innovations and 
Developments in Precision Calculations for the LHC' for hospitality 
during the preparation of this work.


\begin{flushleft}

\end{flushleft}


\clearpage

\begin{appendix}

\section{Shared Memory in Fortran}
\label{app:shm}

IPC shared memory is not natively available in Fortran, but it is not 
difficult to make it available using two small C functions 
\Code{shmalloc} and \Code{shmfree}:
\begin{verbatim}
#include <sys/shm.h>
#include <assert.h>

typedef long long int memindex;
typedef struct { void *addr; int id; } shminfo;

void shmalloc_(shminfo *base, memindex *i, const int *n, const int *size) {
  base->id = shmget(IPC_PRIVATE, *size*(*n + 1) - 1, IPC_CREAT | 0600);
  assert(base->id != -1);
  base->addr = shmat(base->id, NULL, 0);
  assert(base->addr != (void *)-1);
  *i = ((char *)(base->addr + *size - 1) - (char *)base)/(long)*size;
}

void shmfree_(shminfo *base) {
  shmdt(base->addr);
  shmctl(base->id, IPC_RMID, NULL);
}
\end{verbatim}
The function \Code{shmalloc} allocates (suitably aligned) \Code{n} 
elements of size \Code{size} and returns a mock index into \Code{base}, 
through which the memory is addressed in Fortran.  The array \Code{base} 
must be of the desired type and large enough to store the struct 
\Code{shminfo}, \eg two doubles wide.  Be careful to invoke 
\Code{shmfree} after use, for the memory will not automatically be freed 
upon exit but stay allocated until the next reboot (or explicit removal 
with \Code{ipcs}).

The following test program demonstrates how to use \Code{shmalloc} and 
\Code{shmfree}:
\begin{verbatim}
     program test
     implicit none
     integer*8 i
     double precision base(2)

     call shmalloc(base, i, 100, 8)     ! allocate 100 doubles

     base(i) = 1                        ! now use the memory
     ...
     base(i+99) = 100

     call shmfree(base)                 ! don't forget to free it
     end
\end{verbatim}

\end{appendix}


\end{document}